# Compositional Reasoning for Shared-variable Concurrent Programs

Fuyuan Zhang[1], Yongwang Zhao[1], David Sanán[1], Yang Liu[1],
Alwen Tiu[1], Shang-Wei Lin[1], and Jun Sun[2]

[1] School of Computer Science and Engineering, Nanyang Technological University, Singapore
[2] Singapore University of Technology and Design

**Abstract.** Scalable and automatic formal verification for concurrent systems is always demanding. In this paper, we propose a verification framework to support automated compositional reasoning for concurrent programs with shared variables. Our framework models concurrent programs as succinct automata and supports the verification of multiple important properties. Safety verification and simulations of succinct automata are parallel compositional, and safety properties of succinct automata are preserved under refinements. We generate succinct automata from infinite state concurrent programs in an automated manner. Furthermore, we propose the first automated approach to checking rely-guarantee based simulations between infinite state concurrent programs. We have prototyped our algorithms and applied our tool to the verification of multiple refinements.

## 1 Introduction

Automatic verification of concurrent programs is a challenging task. Due to interleaving, the state space of a concurrent program could grow exponentially, which makes it infeasible to directly reason about the global state space. A promising way of conquering the state explosion problem is compositional reasoning [18,33,26,25,36], which aims at breaking the global verification problems into small localized problems. Extensive research [15,14,23,22,19,10] has been conducted on developing rely-guarantee based automatic verification techniques for safety properties of concurrent programs. However, to ensure that safety properties of concurrent programs are preserved after compilation, it is also necessary to show that the checked programs are refined correctly. To the best of our knowledge, all existing approaches to checking rely-guarantee based simulations of concurrent programs [28] are manual.

In this paper, we propose a framework of automated compositional reasoning for shared-variable concurrent programs, which supports both safety verification and refinement checking. In our framework, concurrent programs are modelled as succinct automata, which can be viewed as an extension of program graphs [2]. A succinct automaton consists of both component transitions, specifying behaviors of a local program, and environment transitions, which overapproximate behaviors of other programs in the environment. The idea of integrating these two types of transitions is the key to ensure parallel compositionality. The development of our framework proceeds in the following two directions.

The first direction focuses on parallel compositionalities of safety and simulations of succinct automata, which are very useful in developing compositional proof of global properties. For example, our definition of weak simulations between succinct automata allows compositional reasoning through establishing a local refinement relationship. Let $SA_1$ (resp. $\widehat{SA_1}$) and $SA_2$ (resp. $\widehat{SA_2}$) be two succinct automata and $SA_1 || SA_2$ (resp. $\widehat{SA_1} || \widehat{SA_2}$) be their parallel composition. Since our notion of weak simulation is compositional, we can prove that $SA_1 || SA_2$ weakly refines $\widehat{SA_1} || \widehat{SA_2}$ by proving that $SA_1$ (resp. $SA_2$) weakly refines $\widehat{SA_1}$ (resp. $\widehat{SA_2}$). As safety properties of succinct automata are preserved under refinements, parallel compositionalities of safety and simulations allow us to extend safety properties of high level concurrent programs to low level concurrent programs in compositional ways.

The second direction aims at automating our compositional reasoning techniques. One difficulty of modelling concurrent programs as succinct automata is to find appropriate environment transitions that overapproximate the interleavings between concurrent programs. We show that such environment transitions can be inferred automatically for succinct automata with infinite domains. Moreover, we have developed an SMT-based approach to checking weak simulations between infinite state succinct automata. To the best of our knowledge, we are the first to propose automatic verification of rely-guarantee based simulations for infinite state shared-variable concurrent programs. We have prototyped our tool in F# and verified multiple refinements in automated manner.

Our contributions are fourfold. First, we propose a new formalism, succinct automata, that facilitates automatic verification of multiple properties of shared-variable concurrent programs. Second, we show compositionality results on safety properties and simulations in our framework. Third, we show that succinct automata can be generated automatically from infinite state concurrent programs. Fourth, we provide an SMT-based approach to verifying simulations for infinite state succinct automata.

## 2   Related Work

Extensive research has been conducted on the verification of concurrent programs. Basic approaches to conquering the state explosion problem of concurrent systems include (but not limited to) symbolic model checking [4], partial order reduction [35,32,16], abstraction [17,8,11,9], compositional reasoning [18,33,26,25,31] and symmetry reduction [24,7,13]. The formalism of succinct automata is inspired by rely-guarantee style reasoning [26,25]. We mainly discuss related work on the compositional reasoning of properties considered in this paper.

*Safety Verification.* Our approach to safety verification is closest to thread-modular verification [15], where safety properties are characterized by a set of unsafe states and a global system is safe iff unsafe states are not reachable. In this paper, we focus on invariance properties of succinct automata. Checking strong invariants of succinct automata is dual to verifying whether corresponding sets of unsafe states are reachable. Hence, the approach in [15] can be applied to verify strong invariants of (parallel) succinct automata with finite domains. Work in [23,22,19,10] combined compositional reasoning with abstraction refinement [8]. Moreover, [19,10] allow local variables of different threads to be correlated, which makes their proof rules complete.

*Simulations.* Our work on checking weak simulations is related to previous approaches [3,29,5,28] on compositional reasoning of concurrent programs refinement. In [3,29,5], parallel compositionality is achieved by allowing the environments to have arbitrary behaviors, which is considered too strong in general. Our definition of weak simulations for succinct automata is closely related to and inspired by [28], where a rely-guarantee based simulation, called RGSim, for concurrent programs is proposed. Their compositionality rules for RGSim form the basis of a relational proof method for concurrent programs transformations. Our work differs with theirs mainly in that we aim at developing automatic verification of weak simulations between succinct automata. Also, instead of treating all variables as global variables, we distinguish between local variables and global variables. This greatly reduces the state space of local succinct automata. Compared to [21], which has proposed the first automated proof system for refinement verification of concurrent programs, our approach to refinement checking is more general and is not limited to any specific rules of refinement. Work in [27] proposed an automated refinement checking technique for infinite state CSP programs. Their approach is not developed for shared-variable concurrent programs.

## 3 Succinct Automata

Succinct automata aim to model both local behaviors of a program and its environment in a unified way, and to provide a convenient way to specify useful properties of programs and to support compositional reasoning over them. We distinguish between global variables and local variables when modeling concurrent programs.

### 3.1 Syntax and Semantics

Let *Dom* be a finite or infinite (numeric) domain and $V = \{v_1, ..., v_n\}$ be a finite set of variables ranging over *Dom*. An *atomic predicate* over $V$ is of the form $f(v_1, ..., v_n) \sim b$, where $f : Dom^n \to Dom$ is a function, $\sim \in \{=, <, \leq, >, \geq\}$ and $b \in Dom$. A *predicate* over $V$ is a Boolean combination of atomic predicates over $V$. We write $V'$ for $\{v'_1, ..., v'_n\}$ that refers to variables in $V$ after transitions. Let $\mathcal{F}(V)$ (resp. $\mathcal{F}(V \cup V')$) denote the set of predicates over $V$ (resp. $V \cup V'$). A *valuation* is a function from variables to a domain. Given a valuation $\mathbf{v} : V \to Dom$, we define $n(\mathbf{v}) : V' \to Dom$ as $n(\mathbf{v})(v'_i) = \mathbf{v}(v_i)$ for $v_i \in V$. Given a predicate $\psi \in \mathcal{F}(V_1)$ and a valuation $\mathbf{v} : V_2 \to Dom$, where $V_1 \subseteq V_2$, we write $\psi(\mathbf{v})$ to denote that $\psi$ evaluates to true under the valuation $\mathbf{v}$. We write $\mathbf{Val}_V$ to denote the set of all valuations for variables in $V$.

**Definition 1.** *A Succinct Automaton is a tuple* $SA = (Q, q_0, V, Init, Inv, Env, \Sigma, Edge)$, *where*

- *$Q$ is a finite set of locations and $q_0 \in Q$ is an initial location.*
- *$V = V_G \cup V_L$ and $V_G$ (resp. $V_L$) is a finite set of global (resp. local) variables ranging over Dom, where $V_G \cap V_L = \emptyset$.*
- *$Init \in \mathcal{F}(V)$ defines initial values of variables at $q_0$.*
- *$Inv : Q \to \mathcal{F}(V)$ constrains the values of variables at each location.*
- *$Env : Q \to \mathbf{Val}_{V_G} \times \mathbf{Val}_{V_G}$ specifies environment transitions at each location.*
- *$\Sigma$ is a finite set of action labels which includes the silent action $\tau$.*

- $Edge \subseteq Q \times \Sigma \times \mathcal{F}(V \cup V') \times Q$ *is a finite set of edges specifying component transitions.*

For each location $q \in Q$, transitions specified by $Env(q)$ are made by the environment when $SA$ stays at $q$. In the rest of the paper, we also use predicates or first order formulas to specify $Env(q)$ for convenience. For example, when using $\phi \in \mathcal{F}(V_G \cup V'_G)$ to specify $Env(q)$, $Env(q)$ is defined by $Env(q) = \{(\mathbf{v}_G, \mathbf{v}'_G) \mid \phi(\mathbf{v}_G, n(\mathbf{v}'_G)) \text{ holds}\}$. An edge is of the form $e = (q, \sigma, \mu, q')$, where $\mu$ defines the transition condition and is of the form $\mu := G(V) \wedge \bigwedge_{v'_i \in V'} v'_i = f_i(V)$, where $G(V)$ is a guard for $e$ and $f_i$ is a function $f_i : Dom^n \to Dom$ for $1 \le i \le n$. Action labels in $\Sigma$ are used when we check weak simulations of succinct automata. The main purpose of $Inv$ is to overapproximate reachable states at each control location of a concurrent program. This also facilitates the formalization of the compatibility condition on succinct automata (introduced later). A succinct automaton is *closed* if its environment cannot modify its global variables.

The semantics of succinct automata is defined as a labeled transition system. A *state* of a succinct automaton is a pair $s = (q, \mathbf{v})$ of location $q$ and valuation $\mathbf{v} : V \to Dom$. We denote with $S_{SA}$ the state space of $SA$. A state $(q, \mathbf{v})$ is an initial state iff $q = q_0$ and $Init(\mathbf{v})$ holds. We say that a predicate $\psi$ is satisfied on $(q, \mathbf{v})$ iff $\psi(\mathbf{v})$ holds.

Let $\mathbf{v}_1 : V_1 \to Dom$ and $\mathbf{v}_2 : V_2 \to Dom$ be two valuations such that $V_1 \cap V_2 = \emptyset$. We define $\mathbf{v}_1 \oplus \mathbf{v}_2 : V_1 \cup V_2 \to Dom$ by $\mathbf{v}_1 \oplus \mathbf{v}_2(v) = \mathbf{v}_1(v)$ for $v \in V_1$ and $\mathbf{v}_1 \oplus \mathbf{v}_2(v) = \mathbf{v}_2(v)$ for $v \in V_2$. Let $\mathbf{v}_G : V_G \to Dom$ (resp. $\mathbf{v}_L : V_L \to Dom$) be valuations over global (resp. local) variables. In the rest of the paper, we also use $(q, \mathbf{v}_G \oplus \mathbf{v}_L)$ to represent a state for convenience.

We define two types of transitions, namely *component transitions* and *environment transitions*, for succinct automata. There is a component transition between two states $(q, \mathbf{v}) \xrightarrow{\sigma} (q', \mathbf{v}')$ iff there exists an edge of the form $(q, \sigma, \mu, q') \in Edge$ and $Inv(q)(\mathbf{v}) \wedge \mu(\mathbf{v} \oplus n(\mathbf{v}')) \wedge Inv(q')(\mathbf{v}')$ holds. There is an environment transition between two states $(q, \mathbf{v}) \xrightarrow{env} (q', \mathbf{v}')$ iff $q = q'$, $Inv(q)(\mathbf{v}) \wedge Inv(q)(\mathbf{v}')$ holds, $(\mathbf{v}_G, \mathbf{v}'_G) \in Env(q)$ and $\mathbf{v}_L = \mathbf{v}'_L$, where $\mathbf{v} = \mathbf{v}_G \oplus \mathbf{v}_L$ and $\mathbf{v}' = \mathbf{v}'_G \oplus \mathbf{v}'_L$. Notice that in an environment transition, only values of global variables can be modified and values of local variables remain unchanged.

A *run* of $SA$ is a finite or infinite sequence of environment and component transitions starting from an initial state $(q_0, \mathbf{v}_0)$:

$$(q_0, \mathbf{v}_0) \xrightarrow{env} (q_0, \mathbf{v}'_0) \xrightarrow{\sigma_1} (q_1, \mathbf{v}_1) \xrightarrow{env} (q_1, \mathbf{v}'_1) \xrightarrow{\sigma_2} (q_2, \mathbf{v}_2) \cdots$$

We say that a predicate $\psi$ is satisfied on a run iff it is satisfied on all states on that run.

A finite *local path* of $SA$ is a sequence of edges $\pi = e_1, ..., e_n$, where $e_i = (q_i, \sigma_i, \mu_i, q'_i)$, $e_n = (q_n, \sigma_n, \mu_n, q'_n)$ and $q'_i = q_{i+1}$ for $1 \le i < n$.

We write $(q, \mathbf{v}) \to^* (q', \mathbf{v}')$ if there exists a finite run of $SA$, (consisting of zero or more transitions), from $(q, \mathbf{v})$ to $(q', \mathbf{v}')$ and say that $(q', \mathbf{v}')$ is *reachable* from $(q, \mathbf{v})$. The set of reachable states of $SA$ is the set of states reachable from initial states of $SA$. Regarding environment transitions, we write $(q, \mathbf{v}) \xrightarrow{env^*} (q, \mathbf{v}')$ to denote a finite sequence of environment transitions of $SA$ starting from $(q, \mathbf{v})$ to $(q, \mathbf{v}')$. For component transitions, we write $(q, \mathbf{v}) \xrightarrow{\tau^* \sigma \tau^*} (q', \mathbf{v}')$ to mean that $SA$ has first taken a finite number of silent actions $\tau$, followed by a component transition labelled by an action $\sigma$, and then made another finite number of silent actions.

```
P₁:                              P₂:
while (true) {                   while (true) {
  flag1:=1;                        flag2:=1;
  turn:=2;                         turn:=1;
  await(flag2=0∨turn=1) {         await(flag1=0∨turn=2){
    Critical Section;                Critical Section;
  }                                }
  flag1:=0;                        flag2:=0;
}                                }
```

**Fig. 1.** A Simplified Peterson's Algorithm

*Example 1.* We model a simplified Peterson's algorithm using succinct automata as an example. The pseudo code in Fig. 1 shows a simplified version of Peterson's algorithm with two processes $P_1$ and $P_2$.

In Fig. 2, we model the above two processes as $SA_1 = (Q_1, q_0, V, Init_1, Inv_1, Env_1, \Sigma_1, Edge_1)$ and $SA_2 = (Q_2, p_0, V, Init_2, Inv_2, Env_2, \Sigma_2, Edge_2)$ respectively, where $V = \{flag_1, flag_2, critical_1, critical_2, turn\}$, $\Sigma_1 = \{\tau, c_1\}$ and $\Sigma_2 = \{\tau, c_2\}$. Here, we treat all variables as global variables. The automaton $SA_1$ (resp. $SA_2$) starts at location $q_0$ (resp. $p_0$), where each variable has an initial value of 0, and has five locations $q_0, q_1, q_2, q_3$ and $q_4$ (resp. $p_0, p_1, p_2, p_3$ and $p_4$). Invariants for locations are presented in ovals. Component transitions are represented by solid line arrows, together with the action labels and predicates on them. We omitted the predicates specifying the variables whose values remain unchanged in component transitions. Environment transitions are represented by dashed line arrows and predicates on these arrows specify the binary relations that define environment transitions.

We now briefly explain $SA_1$. At location $q_0$, the environment transition is specified by $\varphi_1 = (flag_1' = flag_1 \wedge critical_1' = critical_1) \wedge (critical_2' = 1 \Rightarrow flag_2' = 1)$, meaning that $SA_2$ never modifies the values of $flag_1$ and $critical_1$ and that if $SA_2$ enters the critical section after the transition, denoted by $critical_2' = 1$, we have $flag_2' = 1$. Then, $SA_1$ takes a silent action to set $flag_1$ to 1, meaning that it wants to enter the critical section, and enters $q_1$. At location $q_1$, the environment transition is specified by $\varphi_2 = (flag_1' = flag_1 \wedge critical_1' = critical_1) \wedge (critical_2' = 1 \Rightarrow (flag_2' = 1 \wedge turn' = 2))$. Compared with $\varphi_1$, we see that if $SA_2$ enters the critical section when $SA_1$ is at $q_1$, $flag_2'$ (resp. $turn'$) must be 1 (resp. 2). This is because $SA_2$ must wait until its turn, denoted by $turn = 2$, to enter the critical section once $SA_1$ has set $flag_1$ to 1. After taking another silent action, $SA_1$ arrives at $q_2$. At location $q_2$, if $flag_2 = 0 \vee turn = 1$, $SA_1$ takes the action $c_1$ and enters the critical section. By entering $q_4$, $SA_1$ leaves the critical section. Finally, $SA_1$ resets $flag_1$ to 0 and comes back to $q_0$.

The environment transitions of $SA_2$ are defined by $\psi_1 = (flag_2' = flag_2 \wedge critical_2' = critical_2) \wedge (critical_1' = 1 \Rightarrow flag_1' = 1)$ and $\psi_2 = (flag_2' = flag_2 \wedge critical_2' = critical_2) \wedge (critical_1' = 1 \Rightarrow (flag_1' = 1 \wedge turn' = 1))$.

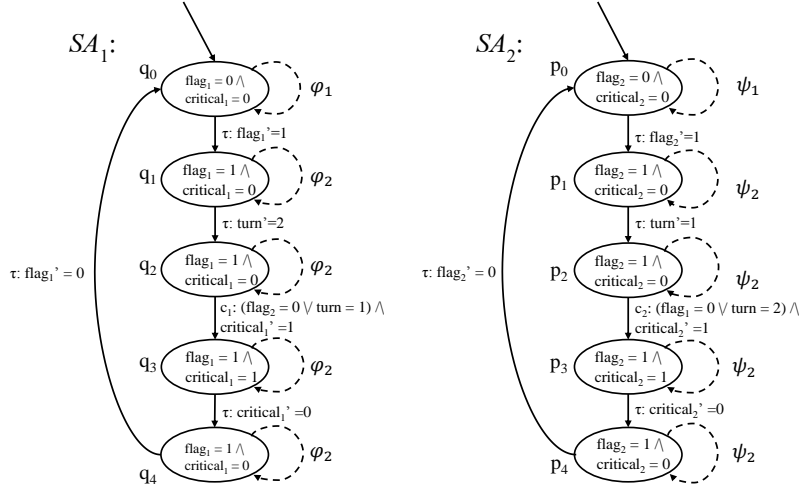

**Fig. 2.** Succinct automata for the Simplified Peterson's algorithm

### 3.2 Parallel Composition

In rely-guarantee reasoning, the guarantee of one thread should imply the rely conditions of other threads. Similarly, we impose a compatibility condition on succinct automata running in parallel. Let $q_1$ (resp. $q_2$) be an arbitrary location in $SA_1$ (resp. $SA_2$). Informally, the compatibility condition guarantees that if $SA_1$ (resp. $SA_2$) makes a component transition from $q_1$ (resp. $q_2$) to $q_1'$ (resp. $q_2'$), $SA_2$ (resp. $SA_1$) can mimic this transition by its environment transitions at $q_2$ (resp. $q_1$). We formalize the *compatibility* condition as follows.

**Definition 2.** $SA_1$ and $SA_2$ are compatible iff for all $(q_1, \boldsymbol{v}_G \oplus \boldsymbol{v}_{L_1}) \in S_{SA_1}$, $(q_2, \boldsymbol{v}_G \oplus \boldsymbol{v}_{L_2}) \in S_{SA_2}$ such that $Inv_1(q_1)(\boldsymbol{v}_G \oplus \boldsymbol{v}_{L_1})$ and $Inv_2(q_2)(\boldsymbol{v}_G \oplus \boldsymbol{v}_{L_2})$, we have

1. *If* $(q_1, \boldsymbol{v}_G \oplus \boldsymbol{v}_{L_1}) \xrightarrow{\sigma_1} (q_1', \boldsymbol{v}_G' \oplus \boldsymbol{v}_{L_1}')$, *then* $(q_2, \boldsymbol{v}_G \oplus \boldsymbol{v}_{L_2}) \xrightarrow{env} (q_2, \boldsymbol{v}_G' \oplus \boldsymbol{v}_{L_2})$.
2. *If* $(q_2, \boldsymbol{v}_G \oplus \boldsymbol{v}_{L_2}) \xrightarrow{\sigma_2} (q_2', \boldsymbol{v}_G' \oplus \boldsymbol{v}_{L_2}')$, *then* $(q_1, \boldsymbol{v}_G \oplus \boldsymbol{v}_{L_1}) \xrightarrow{env} (q_1, \boldsymbol{v}_G' \oplus \boldsymbol{v}_{L_1})$.

Succinct automata running in parallel execute their component transitions in an interleaved manner. The formal definition of parallel composition of compatible succinct automata is defined as follows.

**Definition 3.** Let $SA_1 = (Q_1, q_0^1, V_G \cup V_{L_1}, Init_1, Inv_1, Env_1, \Sigma_1, Edge_1)$ and $SA_2 = (Q_2, q_0^2, V_G \cup V_{L_2}, Init_2, Inv_2, Env_2, \Sigma_2, Edge_2)$ be two compatible succinct automata. The parallel composition of $SA_1$ and $SA_2$ is a succinct automaton $SA_1 \parallel SA_2 = (Q, q_0, V_G \cup V_L, Init, Inv, Env, \Sigma, Edge)$, where

- $Q = Q_1 \times Q_2$, $q_0 = (q_0^1, q_0^2)$, $V_L = V_{L_1} \cup V_{L_2}$ and $\Sigma = \Sigma_1 \cup \Sigma_2$.
- $Init = Init_1 \wedge Init_2$.
- $Inv((q_1, q_2)) = Inv_1(q_1) \wedge Inv_2(q_2)$ for each $q_1 \in Q_1$ and $q_2 \in Q_2$.
- $Env((q_1, q_2)) = Env_1(q_1) \cap Env_2(q_2)$ for each $q_1 \in Q_1$ and $q_2 \in Q_2$.
- $((q_1, q_2), \sigma, \mu, (q_1', q_2')) \in Edge$ iff either:

1. *there exists an edge* $(q_1, \sigma, \mu, q'_1) \in Edge_1$ *and* $q_2 = q'_2$, *or*
2. *there exists an edge* $(q_2, \sigma, \mu, q'_2) \in Edge_2$ *and* $q_1 = q'_1$.

After parallel composition, $SA_1$ and $SA_2$ share a common environment. The environment of $SA_1 \parallel SA_2$ for location $(q_1, q_2)$ is the intersection of the environments of $SA_1$ and $SA_2$ for location $q_1$ and $q_2$ respectively. Intuitively, for each finite run of the parallel composition of two compatible succinct automata, there is a corresponding finite run in each of its components.

## 4 Compositional Reasoning for Succinct Automata

### 4.1 Safety Verification of Succinct Automata

Safety properties require that bad things should not happen. Invariants are a particular kind of safety properties that are useful in specifications. For example, the mutual exclusion property is an invariant which specifies that no more than one thread is in its critical section at any time. We introduce compositional reasoning methods for invariant verification of succinct automata and checking other safety properties can be reduced to invariant verification.

Recall that a predicate $\lambda \in \mathcal{F}(V)$ is an invariant of a transition system $T$ if $\lambda$ is satisfied on all reachable states of $T$. Unlike in a transition system, we have two kinds of transitions, local and environment. The way we treat them leads us to define two types of invariants of succinct automata, strong and weak. When treating both kinds of transitions equally, we reach the notion of strong invariants.

**Definition 4.** *A predicate* $\lambda \in \mathcal{F}(V)$ *is a strong invariant of* $SA$ *if* $\lambda$ *is satisfied on all reachable states of* $SA$.

When focusing on runs of succinct automata where environment transitions preserve $\lambda$, we reach the notion of weak invariants. Here, we say that an environment (resp. a component) transition $(q, \mathbf{v}) \xrightarrow{env} (q', \mathbf{v}')$ (resp. $(q, \mathbf{v}) \xrightarrow{\sigma} (q', \mathbf{v}')$) preserves $\lambda$ if $\lambda(\mathbf{v})$ implies $\lambda(\mathbf{v}')$. The intention of weak invariants is as follows: For a program $T$ modelled as $SA$, if $\lambda$ is a weak invariant of $SA$, then, running in any environment that preserves $\lambda$, $T$ can guarantee that $\lambda$ is preserved in all its local transitions.

**Definition 5.** *A predicate* $\lambda \in \mathcal{F}(V)$ *is a weak invariant of* $SA$ *if* $\lambda$ *is satisfied on all runs of* $SA$ *where environment transitions preserve* $\lambda$.

The notion of weak invariants is more general than strong invariants. In the following, we focus on compositionality of weak invariants. We first impose a noninterference condition on local weak invariants. This condition is to guarantee that local transitions of any component that preserve its own local weak invariant cannot invalidate local weak invariants of other components. Let $\lambda_1$ (resp. $\lambda_2$) be a weak invariant of $SA_1$ (resp. $SA_2$). Formally, we use $noninterfere(\lambda_1, \lambda_2)$ to mean the following condition: $((\lambda_1 \wedge \lambda_2 \wedge \lambda'_1) \Rightarrow \lambda_2[V'_G/V_G]) \wedge ((\lambda_1 \wedge \lambda_2 \wedge \lambda'_2) \Rightarrow \lambda_1[V'_G/V_G])$, where $\lambda'_i$ is derived from $\lambda_i$ by substituting all its variables with corresponding primed variables and $\lambda_i[V'_G/V_G]$ is derived by substituting all global variables in $V_G$ with corresponding primed variables in $V'_G$ for $i = 1, 2$. The parallel compositionality of weak invariants of succinct automata are formalized in the following theorem, which says that local weak invariants satisfied by all the components of the parallel composition of succinct automata guarantee a global weak invariant satisfied by the entire system as long as local weak invariants satisfy the noninterference condition.

**Theorem 1.** *Let $SA_1$ and $SA_2$ be compatible. Assume that $noninterfere(\lambda_1, \lambda_2)$ and $\lambda_1$ (resp. $\lambda_2$) is a weak invariant of $SA_1$ (resp. $SA_2$). We have that $\lambda_1 \wedge \lambda_2$ is a weak invariant of $SA_1 || SA_2$.*

*Example 2.* To show that the simplified Peterson's algorithm in Fig. 1 guarantees mutual exclusion, we check whether $critical_1 = 0 \vee critical_2 = 0$ is a weak invariant of $SA_1 || SA_2$ in Fig. 2. We define $\lambda_1$ and $\lambda_2$ by $\lambda_1 = \lambda_2 = (critical_1 = 0 \vee critical_2 = 0)$. It is easy to verify that $\lambda_1$ (resp. $\lambda_2$) is a weak invariant of $SA_1$ (resp. $SA_2$). Also, it is easy to see that $noninterfere(\lambda_1, \lambda_2)$ holds trivially as $\lambda_1 = \lambda_2$. According to Theorem 1, we know that $critical_1 = 0 \vee critical_2 = 0$ is a weak invariant of $SA_1 || SA_2$, which implies that $P_1$ and $P_2$ in Fig. 1 cannot be in the critical section at the same time.

*Example 3.* We show the correctness of the abstract concurrent GCD programs ($T_1$ and $T_2$) in Fig. 3(a). (The code is taken from [28].) To check that $T_1 || T_2$ really compute the greatest common divisor (gcd) of variables $a$ and $b$, we first model $T_1$ (resp. $T_2$) as $SA_1$ (resp. $SA_2$). The construction of $SA_1$ is shown in Fig. 4 (left), where $\varphi = (a' = a) \wedge (a < b \vee b' = b)$. We omit the construction of $SA_2$ due to space limitation.

For convenience, we introduce two auxiliary variables $A$ and $B$ to $SA_1$ and $SA_2$. The value of $A$ (resp. $B$) equals to the initial value of the input variable $a$ (resp. $b$) and remain unchanged. Let $\lambda_1 = \lambda_2 = (\mathbf{gcd}(a, b) = \mathbf{gcd}(A, B))$, where $\mathbf{gcd}$ is a function that returns the gcd of its input. It is easy to verify that $\lambda_1$ (resp. $\lambda_2$) is a weak invariant of $SA_1$ (resp. $SA_2$). Also, it is easy to see that $noninterfere(\lambda_1, \lambda_2)$ holds. According to Theorem 1, we know that $\mathbf{gcd}(a, b) = \mathbf{gcd}(A, B)$ is a weak invariant of $SA_1 || SA_2$, which implies that $T_1 || T_2$ really compute the gcd of the input values of $a$ and $b$.

## 4.2 Simulations of Succinct Automata

We define weak simulations between succinct automata as follows.

**Definition 6.** *A binary relation $\theta \subseteq S_{SA_1} \times S_{SA_2}$ is a weak simulation for $(SA_1, SA_2)$ w.r.t. a precondition $\kappa \in \mathcal{F}(V_1 \cup V_2)$ and an invariant $\iota \in \mathcal{F}(V_1 \cup V_2)$, denoted by $SA_1 \preceq_{\theta}^{\langle \kappa, \iota \rangle} SA_2$, iff we have the following:*

1. *$\kappa(\mathbf{v}_1, \mathbf{v}_2)$ implies $((q_1, \mathbf{v}_1), (q_2, \mathbf{v}_2)) \in \theta$, where both $q_1$ and $q_2$ are initial.*
2. *$((q_1, \mathbf{v}_1), (q_2, \mathbf{v}_2)) \in \theta$ implies $\iota(\mathbf{v}_1, \mathbf{v}_2)$, $Inv_1(q_1)(\mathbf{v}_1)$, $Inv_2(q_2)(\mathbf{v}_2)$ and the following:*

   a. *if $(q_1, \mathbf{v}_1) \xrightarrow{env} (q_1, \mathbf{v}'_1)$ and $(q_2, \mathbf{v}_2) \xrightarrow{env^*} (q_2, \mathbf{v}'_2)$ and $\iota(\mathbf{v}'_1, \mathbf{v}'_2)$, then we have that $((q_1, \mathbf{v}'_1), (q_2, \mathbf{v}'_2)) \in \theta$.*

   b. *if $(q_1, \mathbf{v}_1) \xrightarrow{\sigma_1} (q'_1, \mathbf{v}'_1)$ and $\sigma_1 \neq \tau$, then there exist $(q'_2, \mathbf{v}'_2) \in S_{SA_2}$ and $\sigma_2 \in \Sigma_2$ such that $\sigma_2 = \sigma_1$, $(q_2, \mathbf{v}_2) \xrightarrow{\tau^* \sigma_2 \tau^*} (q'_2, \mathbf{v}'_2)$ and $((q'_1, \mathbf{v}'_1), (q'_2, \mathbf{v}'_2)) \in \theta$.*

```
T1:                 T2:                 T1':                T2':
m:=0;               n:=0;               m:=0;               n:=0;
while (m=0) {       while (n=0) {       while (m=0) {       while (n=0) {
   atomic {            atomic {            x1:=a;              y1:=a;
     if (a=b)            if (a=b)          x2:=b;              y2:=b;
       m:=1;              n:=1;            if (x1=x2)         if (y1=y2)
     if (a>b)           if (a<b)             m:=1;              n:=1;
       a:=a-b;            b:=b-a;         if (x1>x2)         if (y1<y2)
   }                  }                     a:=x1-x2;          b:=y2-y1;
}                   }                   }                   }
```

(a) Abstract GCD Programs                   (b) Concrete GCD Programs

**Fig. 3.** Concurrent GCD Programs

    **c.** *if* $(q_1, \boldsymbol{v}_1) \xrightarrow{\tau} (q_1', \boldsymbol{v}_1')$, *then there exists* $(q_2', \boldsymbol{v}_2') \in S_{SA_2}$ *such that* $(q_2, \boldsymbol{v}_2) \xrightarrow{\tau^*} (q_2', \boldsymbol{v}_2')$ *and* $((q_1', \boldsymbol{v}_1'), (q_2', \boldsymbol{v}_2')) \in \theta$.

Conditions 2.**b** and 2.**c** constrain local behaviors of $SA_1$ and $SA_2$ and are similar to standard notions of weak simulations [30]. Condition 2.**a** constrains the environments of the two succinct automata and requires that the weak simulation should not be affected by the environments as long as the valuations of variables in $V_1$ and $V_2$ are related by $\iota$. Note that if merely we were to require that an environment transition from $q_1$ is simulated by zero or more environment transitions from $q_2$, the resulting simulation relation would not be compositional under parallel composition. Our way of dealing with environments in defining simulation or bi-simulation relations is not without precedent. For example, in process calculi, e.g., higher-order calculi [34] or cryptographic calculi [1], environments are treated separately from local transitions, and one typically requires certain relations to hold between the environments, e.g., as in the relation $\iota$ we have above. Condition 2.**a** is the key for compositionality in our notion of weak simulation.

Given $\kappa$ and $\iota$, we say that $SA_1$ is *weakly simulated by* $SA_2$ (or $SA_1$ *weakly refines* $SA_2$) *with respect to* $\kappa$ *and* $\iota$, denoted by $SA_1 \preceq^{(\kappa, \iota)} SA_2$, if there exists a weak simulation $\theta$ such that $SA_1 \preceq_\theta^{(\kappa, \iota)} SA_2$. We say that $SA_1$ is *weakly simulated by* $SA_2$, denoted by $SA_1 \preceq SA_2$, if there exist $\kappa$ and $\iota$ such that $SA_1 \preceq^{(\kappa, \iota)} SA_2$. The relation $\preceq$ on succinct automata is reflexive but not transitive. However, the relation $\preceq$ on closed succinct automata is transitive. This allows us to chain together two refinement steps when reasoning about simulations between closed succinct automata.

**Theorem 2.** *The relation* $\preceq$ *on closed succinct automata forms a pre-order.*

For succinct automata that are not closed, we can still chain together successive refinement steps if the environment transitions of related succinct automata satisfy a certain condition. We formalize this in the following theorem.

**Theorem 3.** *Assume that* $SA_1 \preceq_{\theta_1}^{(\kappa_1, \iota_1)} SA_2$ *and* $SA_2 \preceq_{\theta_2}^{(\kappa_2, \iota_2)} SA_3$. *Let* $\kappa, \iota \in \mathcal{F}(V_1 \cup V_3)$ *be predicates such that* $\kappa(\boldsymbol{v}_1, \boldsymbol{v}_3)$ *(resp.* $\iota(\boldsymbol{v}_1, \boldsymbol{v}_3)$) *holds iff there exists* $\boldsymbol{v}_2$ *such that* $\kappa_1(\boldsymbol{v}_1, \boldsymbol{v}_2) \wedge \kappa_2(\boldsymbol{v}_2, \boldsymbol{v}_3)$ *(resp.* $\iota_1(\boldsymbol{v}_1, \boldsymbol{v}_2) \wedge \iota_2(\boldsymbol{v}_2, \boldsymbol{v}_3)$) *holds. We have that* $SA_1 \preceq_{\theta_2 \circ \theta_1}^{(\kappa, \iota)} SA_3$ *if the following holds: Assume that* $(q_1, \boldsymbol{v}_1) \xrightarrow{env} (q_1, \boldsymbol{v}_1')$, $(q_3, \boldsymbol{v}_3) \xrightarrow{env^*} (q_3, \boldsymbol{v}_3')$, $\iota(\mathbf{v}_1, \mathbf{v}_3)$ and $\iota(\mathbf{v}_1', \mathbf{v}_3')$. For any $\mathbf{v}_2$ such that $\iota_1(\mathbf{v}_1, \mathbf{v}_2) \wedge \iota_2(\mathbf{v}_2, \mathbf{v}_3)$ and for all $q_2 \in Q_2$, there exists $\mathbf{v}_2'$ such that $(q_2, \mathbf{v}_2) \xrightarrow{env} (q_2, \mathbf{v}_2')$ and $\iota_1(\mathbf{v}_1', \mathbf{v}_2') \wedge \iota_2(\mathbf{v}_2', \mathbf{v}_3')$.

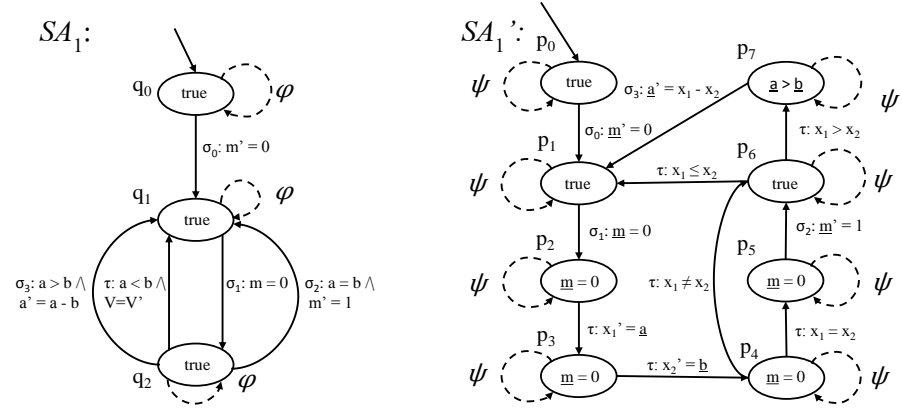

**Fig. 4.** Succinct automata for Concurrent GCD

Given $\theta_1 \subseteq S_{SA_1} \times S_{\widehat{SA_1}}$ and $\theta_2 \subseteq S_{SA_2} \times S_{\widehat{SA_2}}$, we define $\theta_1 \otimes \theta_2 \subseteq S_{SA_1||SA_2} \times S_{\widehat{SA_1}||\widehat{SA_2}}$ as follows: $(((q_1, q_2), \mathbf{v}), ((\widehat{q_1}, \widehat{q_2}), \widehat{\mathbf{v}})) \in \theta_1 \otimes \theta_2$ iff $((q_1, \mathbf{v}_G \oplus \mathbf{v}_{L_1}), (\widehat{q_1}, \widehat{\mathbf{v}_G} \oplus \widehat{\mathbf{v}_{L_1}})) \in \theta_1$ and $((q_2, \mathbf{v}_G \oplus \mathbf{v}_{L_2}), (\widehat{q_2}, \widehat{\mathbf{v}_G} \oplus \widehat{\mathbf{v}_{L_2}})) \in \theta_2$, where $\mathbf{v} = \mathbf{v}_G \oplus \mathbf{v}_{L_1} \oplus \mathbf{v}_{L_2}$ and $\widehat{\mathbf{v}} = \widehat{\mathbf{v}_G} \oplus \widehat{\mathbf{v}_{L_1}} \oplus \widehat{\mathbf{v}_{L_2}}$.

To ensure compositionality of weak simulations, we also impose a noninterference condition on $\iota_1$ and $\iota_2$ here. We reuse $noninterfere(\iota_1, \iota_2)$ to denote the following condition: $((\iota_1 \wedge \iota_2 \wedge \iota_1') \Rightarrow \iota_2[V_G'/V_G][\widehat{V_G}'/\widehat{V_G}]) \wedge ((\iota_1 \wedge \iota_2 \wedge \iota_2') \Rightarrow \iota_1[V_G'/V_G][\widehat{V_G}'/\widehat{V_G}])$. The following theorem shows that weak simulations of succinct automata are preserved under parallel composition.

**Theorem 4.** *Assume that $SA_1$ (resp. $\widehat{SA_1}$) and $SA_2$ (resp. $\widehat{SA_2}$) are compatible and that $noninterfere(\iota_1, \iota_2)$. We have that $SA_1 \preceq_{\theta_1}^{(\kappa_1, \iota_1)} \widehat{SA_1}$ and $SA_2 \preceq_{\theta_2}^{(\kappa_2, \iota_2)} \widehat{SA_2}$ implies $SA_1||SA_2 \preceq_{\theta_1 \otimes \theta_2}^{(\kappa_1 \wedge \kappa_2, \iota_1 \wedge \iota_2)} \widehat{SA_1}||\widehat{SA_2}$.*

*Example 4.* We show that the abstract concurrent GCD programs ($T_1$ and $T_2$) in Fig. 3(a) are refined by the concrete GCD programs ($T_1'$ and $T_2'$) in Fig. 3(b). The bodies of the while loops in $T_1$ and $T_2$ are executed atomically and are refined to corresponding code in $T_1'$ and $T_2'$ to allow interleaving.

In Fig. 4, we model thread $T_1$ (resp. $T_1'$) as $SA_1$ (resp. $SA_1'$), where $\varphi = (a' = a) \wedge (a < b \vee b' = b)$ and $\psi = (\underline{a}' = \underline{a}) \wedge (\underline{a} < \underline{b} \vee \underline{b}' = \underline{b})$. Let $\kappa_1$ and $\iota_1$ be defined by $\kappa_1 = (\underline{a} = a \wedge \underline{b} = b \wedge \underline{m} = m)$ and $\iota_1 = (\underline{a} = a \wedge \underline{b} = b \wedge \underline{m} = m)$. In our experiment, using the verification tool we have implemented, we have verified that $SA_1' \preceq_{\theta_1}^{(\kappa_1, \iota_1)} SA_1$ holds for some $\theta_1$. Similarly, we have modeled $T_2$ (resp. $T_2'$) as $SA_2$ (resp. $SA_2'$) and checked in our experiment that $SA_2' \preceq_{\theta_2}^{(\kappa_2, \iota_2)} SA_2$ holds for some $\theta_2$. By Theorem 4, we have that $SA_1'||SA_2' \preceq_{\theta_1 \otimes \theta_2}^{(\kappa_1 \wedge \kappa_2, \iota_1 \wedge \iota_2)} SA_1||SA_2$.

### 4.3 Safety Property Preservation under Refinement

It is obvious that strong invariants are preserved under refinements. We show in the following that weak invariants of succinct automata are also preserved under refinements.

We write $\mathcal{WS}(\theta, Env_1, Env_2)$ to mean that: if $((q_1, \mathbf{v}_1), (q_2, \mathbf{v}_2)) \in \theta$ and $(q_1, \mathbf{v}_1) \xrightarrow{env} (q_1, \mathbf{v}_1')$, there exists $(q_2, \mathbf{v}_2') \in S_{SA_2}$ such that $(q_2, \mathbf{v}_2) \xrightarrow{env^*} (q_2, \mathbf{v}_2')$ and $((q_1, \mathbf{v}_1'), (q_2, \mathbf{v}_2')) \in \theta$. If $\mathcal{WS}(\theta, Env_1, Env_2)$ holds, for each run in $SA_1$, we can construct a corresponding run in $SA_2$ such that the two runs are related by $\theta$. Thus, we have the following lemma that links reachability and weak simulations.

**Lemma 1.** *Assume that $SA_1 \preceq_\theta^{(\kappa, \iota)} SA_2$ holds for some $\theta, \kappa$ and $\iota$, where $\mathcal{WS}(\theta, Env_1, Env_2)$ holds. For all states $(q_1, \mathbf{v}_1) \in S_{SA_1}$ and $(q_2, \mathbf{v}_2) \in S_{SA_2}$ such that $((q_1, \mathbf{v}_1), (q_2, \mathbf{v}_2)) \in \theta$, if $(q_1, \mathbf{v}_1) \rightarrow^* (q_1', \mathbf{v}_1')$ for some $(q_1', \mathbf{v}_1') \in S_{SA_1}$, there exists $(q_2', \mathbf{v}_2') \in S_{SA_2}$ such that $(q_2, \mathbf{v}_2) \rightarrow^* (q_2', \mathbf{v}_2')$ and $((q_1', \mathbf{v}_1'), (q_2', \mathbf{v}_2')) \in \theta$.*

As invariants verification can be reduced to reachability problems, we can prove by contradiction that the following theorem holds.

**Theorem 5.** *Assume that $SA_1 \preceq_\theta^{(\kappa, \iota)} SA_2$ holds for some $\theta, \kappa$ and $\iota$, where $\mathcal{WS}(\theta, Env_1, Env_2)$ holds, and for each initial state $(q_1, \mathbf{v}_1) \in S_{SA_1}$, there exists an initial state $(q_2, \mathbf{v}_2) \in S_{SA_2}$ such that $\kappa(\mathbf{v}_1, \mathbf{v}_2)$. Let $\lambda_1 \in \mathcal{F}(V_1)$ and $\lambda_2 \in \mathcal{F}(V_2)$ be two predicates such that $\neg\lambda_1(\mathbf{v}_1) \land \iota(\mathbf{v}_1, \mathbf{v}_2)$ implies $\neg\lambda_2(\mathbf{v}_2)$. If $\lambda_2$ is a weak invariant of $SA_2$, then $\lambda_1$ is a weak invariant of $SA_1$.*

*Example 5.* We give a short example to show that $\mathbf{gcd}(a, b) = \mathbf{gcd}(A, B)$ is a weak invariant of the concrete GCD programs, which implies that the concrete GCD programs also compute the gcd of the input variables. First, we know from Example 3 that $\mathbf{gcd}(a, b) = \mathbf{gcd}(A, B)$ is a weak invariant of the abstract GCD programs. Second, we know from Example 4 that the concrete GCD programs refine the abstract GCD programs. Hence, from Theorem 5, we can prove that $\mathbf{gcd}(a, b) = \mathbf{gcd}(A, B)$ is also a weak invariant of the concrete GCD programs.

## 5 Automatic Verification of Succinct Automata

We focus on two aspects of automated verification of succinct automata: generation of succinct automata from infinite state concurrent programs and refinement checking between infinite state succinct automata. We prototyped our tool in the functional programming language F# in over 3700 lines of code and used Z3 [12] in our implementation. We applied our tool to check multiple weak simulations between concurrent C programs. Experimental results are included in the appendix.

### 5.1 Generation of Succinct Automata

The hardest part of generating succinct automata from infinite state concurrent programs is to construct their invariant components and environment components. Intuitively, invariant components overapproximate reachable states at control locations of concurrent programs and environment components abstract the transitions of other programs in the environment. To construct these components, we perform separate forward reachability analysis for each concurrent program on abstract domains, and for component transitions of a concurrent program that modify global variables, corresponding environment transitions are generated for other concurrent programs in the environment. We present our algorithm for generating succinct automata in Algorithm 1.

The main function in Algorithm 1 is $Generate\text{-}SAs$. Given two concurrent programs $T_1$ and $T_2$, it first constructs two intermediate automata $SA_1$ and $SA_2$, where $Inv_1, Env_1, Inv_2$ and $Env_2$ are not specified. At this step, $SA_1$ and $SA_2$ are essentially the program graphs of $T_1$ and $T_2$. Then, it initializes $Inv_i$ and $EnvSet_i$. Here, $EnvSet_i$ is used to keep track of changes of global variables made by $SA_j$, where $i \neq j$ and $i, j = 1, 2$. After that, it starts fixed-point iterations (Line 21-26) to overapproximate reachable states at each location by calling function $Reach$ and generate corresponding environment transitions by calling function $GenEnvTrans$. After the least fixed points are reached, it constructs $Env_i$ from $EnvSet_i$. If the relation specified by $EnvSet_i$ is not reflexive, we explicitly add $V_G = V'_G$ to make $Env_i$ reflexive.

Function $Reach$ performs the forward reachability analysis for $SA_i$, where $EnvSet_i$ specifies the environment transitions of $SA_i$. Function $Post_{Comp}(Inv_i(q), \mu)$ (Line 4) calculates a predicate that overapproximates states reachable from $Inv_i(q)$ by executing a component transition whose transition condition is $\mu$. Function $Post_{Env}(Inv_i(q), EnvSet_i)$ (Line 6) calculates a predicate that overapproximates states reachable from $Inv_i(q)$ by executing environment transitions specified by $EnvSet_i$.

Function $GenEnvTrans$ takes $Inv_i$ and $Edge_i$ of $SA_i$ and generates environment transitions for $SA_j$, where $i \neq j$ and $i, j = 1, 2$. For each edge $(q, \sigma, \mu, q') \in Edge_i$ that modifies global variables, we generate a corresponding pair $(Inv_i(q), \mu)$ (Line 14) to be used to specify environment transitions of $SA_j$. Function $GenEnvTrans$ is the key to guarantee the compatibility of $SA_1$ and $SA_2$.

We have the following theorem that guarantees the compatibility of $SA_1$ and $SA_2$.

**Theorem 6.** *$SA_1$ and $SA_2$ generated by Algorithm 1 are compatible.*

From Theorem 6, it's easy to prove by contradiction that $SA_1||SA_2$ overapproximates $T_1||T_2$, which means for each execution trace of $T_1||T_2$, there is a corresponding run of $SA_1||SA_2$.

In our prototype, the abstract domain we use is Boxes [20] and an element on the Boxes domain is implemented as a corresponding Linear Decision Diagram (LDD) [6]. To guarantee the termination of the iteration in $Reach$, we used widening techniques [11] for the Boxes domain, which is not listed in Algorithm 1 due to space limitation. On the other hand, we point out here that Algorithm 1 is a general algorithm that can be implemented on top of other abstract domains and the correctness of Theorem 6 is independent of the abstract domains underlying Algorithm 1.

### 5.2 Refinement Checking between Succinct Automata

We propose an SMT-based approach (Algorithm 2) to checking weak simulations between infinite state succinct automata. One difficulty in developing an SMT-based approach here comes from Condition 2.**a** in Definition 6, because environment transitions of the abstract succinct automata can be executed arbitrary finite number of times.

---

**Algorithm 1:** Generating Succinct Automata from Concurrent Programs

---

**Input:** Concurrent programs $T_1$ and $T_2$.
**Output:** Compatible $SA_1$ and $SA_2$ that models $T_1$ and $T_2$.

1   **Function** $Reach\,(SA_i, EnvSet_i)$ **is**
2     **repeat**
3       **foreach** $(q, \sigma, \mu, q') \in Edge_i$ **do**
4         $Inv_i(q') := Inv_i(q') \vee Post_{Comp}(Inv_i(q), \mu)$
5       **foreach** $q \in Q_i$ **do**
6         $Inv_i(q) := Inv_i(q) \vee Post_{Env}(Inv_i(q), EnvSet_i)$
7     **until** *No more reachable states are added to $Inv_i(q)$ for all $q \in Q_i$*
8     **return** $Inv_i$

10   **Function** $GenEnvTrans\,(Inv_i, Edge_i, j)$ **is**
11     $EnvSet_j := \emptyset$
12     **foreach** $(q, \sigma, \mu, q') \in Edge_i$ **do**
13       **if** *$\mu$ modifies global variables* **then**
14         $EnvSet_j := EnvSet_j \cup (Inv_i(q), \mu)$
15     **return** $EnvSet_j$

17   **Function** `Generate-SAs`$(T_1, T_2)$ **is**
18     Construct intermediate succinct automata $SA_1$ and $SA_2$
19     $Inv_i(q) := false$ for all $q \in Q_i$ and $i = 1, 2$
20     $EnvSet_i = \emptyset$ for $i = 1, 2$
21     **repeat**
22       $Inv_1 := Reach(SA_1, EnvSet_1)$
23       $Inv_2 := Reach(SA_2, EnvSet_2)$
24       $EnvSet_1 := GenEnvTrans(Inv_2, Edge_2, 1)$
25       $EnvSet_2 := GenEnvTrans(Inv_1, Edge_1, 2)$
26     **until** *Least Fixed Points are Reached*
27     Construct $Env_i$ from $EnvSet_i$ and make $Env_i$ reflexive for $i = 1, 2$
28     **return** $Inv_1, Inv_2, Env_1, Env_2$

---

However, we have noticed in practice that the length of local paths of succinct automata whose action labels are of the form $\tau^* \sigma \tau^*$ or $\tau^*$ are usually bounded. Hence, in Algorithm 2, we only specify the execution of environment transitions of the abstract succinct automata up to a bound $k$, which is precalculated by our prototyped tool.

Proving $SA_1 \preceq^{(\kappa, \iota)} SA_2$ amounts to showing the existence of a simulation relation $\theta$ such that $SA_1 \preceq_\theta^{(\kappa, \iota)} SA_2$. We define first order formulas $\Psi_{(q_1, q_2)}$ over $V_1 \cup V_2$ for a set of pairs of locations $(q_1, q_2) \in Q_1 \times Q_2$. The intention is that when our algorithm terminates, we can construct a relation $\theta = \{((q_1, \mathbf{v}_1), (q_2, \mathbf{v}_2)) \mid \Psi_{(q_1, q_2)}(\mathbf{v}_1, \mathbf{v}_2) \text{ holds}\}$ such that $\theta$ satisfies Condition 2 in Definition 6.

Our algorithm follows the basic fixed point iteration method. The main function in Algorithm 2 is $Check\text{-}Weak\text{-}Simulation$. It first computes a set $\Theta$ that contains all

---

**Algorithm 2:** An Algorithm to Check Weak Simulations of Succinct Automata

---

**Input:** $SA_1$ and $SA_2$ and parameters $\kappa$ and $\iota$.

**Output:** If the algorithm return *Yes*, $SA_1 \preceq^{\langle \kappa, \iota \rangle} SA_2$ holds. If the algorithm returns *No*, $SA_1 \preceq^{\langle \kappa, \iota \rangle} SA_2$ does not hold.

**1 Function** $GenConstraints\,(SA_1, SA_2, \Theta, \iota)$ **is**

**2**    **foreach** $(q_1, q_2) \in \Theta$ **do**

**3**      $C_1 := \Psi_{(q_1, q_2)} \Rightarrow ((\Phi_{Env_1(q_1)} \wedge \iota[V_1'/V_1]) \Rightarrow \Psi_{(q_1, q_2)}[V_1'/V_1])$

**4**      $constraints := constraints \cup \{\neg C_1\}$

**5**      **foreach** $1 \leq j \leq k$ **do**

**6**        $C_2^j := \Psi_{(q_1, q_2)} \wedge ((\Phi_{Env_1(q_1)} \wedge \Phi_{Env_2(q_2)}^j \wedge$

**7**          $\iota[V_1'/V_1][V_2^j/V_2]) \Rightarrow \Psi_{(q_1, q_2)}[V_1'/V_1][V_2^j/V_2])$

**8**        $constraints := constraints \cup \{\neg C_2^j\}$

**9**      **foreach** $e = (q_1, \sigma, \mu, q_1') \in Edge_1$ **do**

**10**        $C_3 := \Psi_{(q_1, q_2)} \Rightarrow (G \Rightarrow (WP(e', \bigvee_{\pi \in \Pi_\sigma(q_2)} WP(\pi, \Psi_{(q_1', q_2')}))))$

**11**          where $G$ is the guard of $e$, $e'$ is derived from $e$ by substituting its guard

**12**          with $True$, and $\pi$ ends at location $q_2'$

**13**        $constraints := constraints \cup \{\neg C_3\}$

**14**    **return** $constraints$

**15**

**16 Function** $UpdatePsi\,(constraints, V_1, V_2, \Theta)$ **is**

**17**    **foreach** $(q_1, q_2) \in \Theta$ **do**

**18**      $\Psi'_{(q_1, q_2)} := \Psi_{(q_1, q_2)}$

**19**    **foreach** $\neg(\Psi_{(q_1, q_2)} \Rightarrow \Phi) \in constraints$ **do**

**20**      **if** $\neg(\Psi_{(q_1, q_2)} \Rightarrow \Phi)$ *is satisfiable* **then**

**21**        **if** $\neg(\Psi_{(q_1, q_2)} \Rightarrow \Phi)$ *is a type 1 constraint* **then**

**22**          $\Psi'_{(q_1, q_2)} := \Psi'_{(q_1, q_2)} \wedge \forall V\, \Phi$

**23**          where $V = FreeVar(\Phi) \backslash (V_1 \cup V_2)$

**24**        **if** $\neg(\Psi_{(q_1, q_2)} \Rightarrow \Phi)$ *is a type 2 constraint* **then**

**25**          $\Psi'_{(q_1, q_2)} := \Psi'_{(q_1, q_2)} \wedge \Phi$

**26**    **foreach** $(q_1, q_2) \in \Theta$ **do**

**27**      $\Psi_{(q_1, q_2)} := \Psi'_{(q_1, q_2)}$

**28**    **if** *none of the constraints are satisfiable* **then**

**29**      **return** *Fixed Point Reached*

**30**    **else**

**31**      **return** *Continue Iteration*

**32**

**33 Function** `Check-Weak-Simulation`$(SA_1, SA_2, \kappa, \iota)$ **is**

**34**    $\Theta := GenPairs(\{(q_{init_1}, q_{init_2})\})$

**35**    **foreach** $(q_1, q_2) \in \Theta$ **do**

**36**      $\Psi_{(q_1, q_2)} := \iota \wedge Inv_1(q_1) \wedge Inv_2(q_2)$

**37**    $constraints := \emptyset$

**38**    **repeat**

**39**      $constraints := GenConstraints(SA_1, SA_2, \Theta, \iota)$

**40**      $result := UpdatePsi(constraints, V_1, V_2, \Theta)$

**41**    **until** $result = Fixed\ Point\ Reached$

**42**    **if** $\kappa \Rightarrow \Psi_{(q_{init1}, q_{init2})}$ *is valid* **then**

**43**      **return** *Yes*

**44**    **else**

**45**      **return** *No*

the pairs $(q_1, q_2)$ for which we need to define constraints. Then, it defines the initial value of $\Psi_{(q_1,q_2)}$ for each $(q_1, q_2) \in \Theta$. In each fixed point iteration (Line 38-41), we first generate constraints for each $(q_1, q_2) \in \Theta$ that specify Condition 2 of Definition 6 by calling function $GenConstraints$. Then, we refine the value of $\Psi_{(q_1,q_2)}$ through function $UpdatePsi$ according to the satisfiability of the constraints generated for $(q_1, q_2)$. When the greatest fixed point is reached, it is guaranteed that Condition 2 of Definition 6 is satisfied. Finally, we check whether Condition 1 of Definition 6 is also satisfied (Line 42).

Due to space limitation, we omit the pseudo code for the function $GenPairs$ (called in Line 34) and explain it briefly as follows. Let $\Pi_\sigma(q)$ denote the set of finite local paths $\pi$ such that $\pi$ starts from $q$ and the action labels along $\pi$ are of the form $\tau^* \sigma \tau^*$ (resp. $\tau^*$), when $\sigma \neq \tau$ (resp. $\sigma = \tau$). $GenPairs$ is a recursive function which takes a set $\Theta$ of pairs of locations as input and returns another set of pairs of locations. Let $\Theta'$ be an empty set. First, for each $(q_1, q_2) \in \Theta$, it adds to $\Theta'$ the set of $(q_1', q_2')$ such that there exists an edge $(q_1, \sigma, \mu, q_1')$ and a path $\pi \in \Pi_\sigma(q_2)$ that ends in $q_2'$. Then, $GenPairs$ makes a recursive call $GenPairs(\Theta' \backslash \Theta)$ and returns $\Theta \cup GenPairs(\Theta' \backslash \Theta)$.

Function $GenConstraints$ generates following constraints $\neg C_1, \neg C_2^1, ..., \neg C_2^k$ and $\neg C_3$ for each $(q_1, q_2) \in \Theta$. Formulas $C_1$ and $C_2^j$ (Line 3 and 6-7) are used to specify Condition 2.**a** in Definition 6, where $\Phi_{Env_1(q_1)}$ is a predicate that specifies the execution of environment transitions $Env_1(q_1)$ once and $\Phi_{Env_2(q_2)^j}$ is a predicate specifying the execution of environment transitions $Env_2(q_2)$ for $j$ steps. In Line 7, we write $V_2^j$ to mean $\{v_1^j, ..., v_n^j\}$ for $V_2 = \{v_1, ..., v_n\}$. Formula $C_3$ (Line 10) specifies Condition 2.**b** and 2.**c**. We use $WP(e, \Psi)$ (resp. $WP(\pi, \Psi)$) to denote the weakest precondition such that $\Psi$ holds after taking a component transition (resp. a sequence of component transitions) by executing $e$ (resp. $\pi$).

Function $UpdatePsi$ checks the satisfiability of all the constraints generated by $GenConstraints$. If a constraint $\neg(\Psi_{(q_1,q_2)} \Rightarrow \Phi)$ is satisfiable, $\Psi_{(q_1,q_2)}$ fails to satisfy Condition 2 in Definition 6. In this case, we strengthen $\Psi_{(q_1,q_2)}$ in Line 21-25 depending on the type of the constraint. Here, type 1 (resp. type 2) constraints refer to those of the form $\neg C_1$ and $\neg C_2^j$ (resp. $\neg C_3$) generated by $GenConstraints$.

# 6 Conclusions and Future Work

In this paper, we have laid the theoretical underpinning for succinct automata, which is a formalism for formal verification of shared-variable concurrent programs. In our framework, safety verification and simulations of concurrent programs are parallel compositional and algorithmic. Succinct automata-based approaches can be applied to extend safety verification of concurrent programs from the source code level down to the binary level in a compositional way.

At the current stage, our prototype is able to verify refinements between concurrent C programs. Compared with manual proofs, our automated verification technique saves considerable time. In our future work, we will study how to generate succinct automata from assembly code and further develop our tool so that it can verify refinements between concurrent C programs and assembly code.

## A   Experimental Results

We applied our tool to check multiple weak simulations between concurrent programs. We have conducted all our experiments on a desktop with an Intel 3.50 GHz twelve-core processor, 16 GB memory, running Windows 7. The experimental results are summarized in Table 1, where the second column refers to the time for verification and the third column indicates whether we found the simulation relations in the experiment.

**Table 1.** Refinement Checking between Concurrent Programs

| Subject | Time: sec | Results |
|---|---|---|
| Concurrent GCD | 3.299 ($t1$) | Yes ($t1$) |
|  | 2.709 ($t2$) | Yes ($t2$) |
| Peterson's Algorithm | 0.937 | No |
| Invariant Hoisting | 1.661 | Yes |
| Strength Reduction | 156.518 | Yes |
| Induction Variable Elimination | 218.695 | Yes |
| Dead Code Elimination | 14.413 | No |
| Constant Folding and Constant Propagation | 5.481 | Yes |
| If Optimization (1) | 49.015 | Yes |
| If Optimization (2) | 48.997 | Yes |
| Cross Jumping | 415.490 | Yes |
| Branch Elimination | 19.412 | Yes |
| Loop Unrolling | 10.412 | Yes |

### A.1   Concurrent GCD and Peterson's Algorithm

In the experiment of Concurrent GCD, we focused on the refinement problem in Example 4 and verified that the concrete GCD programs in Fig. 3(b) do refine the abstract concurrent GCD programs in Fig. 3(a). As there are two threads in GCD programs, experimental results annotated with $t1$ (resp. $t2$) are related to checking refinements of the first (resp. second) threads.

In the experiment of Peterson's algorithm, we switched the order of the first two instructions in the while loop of the simplified Peterson's algorithm in Fig. 1 and have found that the modified loop body does not refine the original one. Actually, the modified Peterson's algorithm does not guarantee mutual exclusion. This is consistent with our theorems because if the modified Peterson's algorithm refines the one in Fig. 1, it should also guarantee mutual exclusion. The experimental results in the fourth row are about checking whether the modified loop body of $P_1$ in Fig. 1 refines the original one.

### A.2   Compiler Optimizations

In the rest of the experiments, we checked the correctness of several compiler optimizations on concurrent programs. The compiler optimizations involved are often used for sequential programs in practice. In concurrent settings, due to unintended behaviors of the environments, these optimizations are not always correct. Actually, they are correct only when the environments are reasonably constrained. We summarize the setting of our experiments as follows.

For each category of compiler optimizations in our experiment, a source program $C$ is transformed to a target program $C'$ by compiler optimizations, and the assumptions on the environments of $C$ and $C'$ are made. We construct a program $C_e$ serving as a concurrent program running in the environments and verify whether $C'||C_e$ refines $C||C_e$. Program $C_e$ is constructed such that it follows our assumptions on the environments. Our tool generates $SA||SA_e$ (resp. $SA'||SA'_e$) from $C||C_e$ (resp. $C'||C_e$) automatically, and we provide corresponding parameters $\kappa$ and $\iota$ to our tool. These predicates mainly specify that common variables in $SA||SA_e$ and $SA'||SA'_e$ should have the same values. Our tool verified the refinements between $SA'||SA'_e$ and $SA||SA_e$ successfully.

For each category of compiler optimization, we give the example code and our assumptions on environment transitions.

***Invariant Hoisting*** The source program $C_1$ in Fig. 5 is transformed to the target program $C_2$ in Fig. 5 by invariant hoisting. Both $C_1$ and $C_2$ are supposed to run in an environment that does not modify the values of $x$ and $t$. We constructed a program $C_e$ in Fig. 5 and verified that $C_2||C_e$ refines $C_1||C_e$.

| Source Program $C_1$ | Target Program $C_2$ | Program $C_e$ |
|---|---|---|
| ```while(i<n) {``` | ```t:=x+5;``` | ```n:=n+1;``` |
| ```    t:=x+5;``` | ```while(i<n) {``` | ```k:=5*x+1;``` |
| ```    i:=2*i+t;``` | ```    i:=2*i+t;``` | ```i:=i+3;``` |
| ```}``` | ```}``` | |

**Fig. 5.** Invariant Hoisting

***Strength Reduction*** The source program $C_1$ in Fig. 6 is transformed to a target program $C_2$ in Fig. 6 by strength reduction. We assume that the environment of $C_1$ does not modify $i$ and the environment of $C_2$ does not modify $i$ and $k$. We constructed a program $C_e$ in Fig. 6 and verified that $C_2||C_e$ refines $C_1||C_e$.

| Source Program $C_1$ | Target Program $C_2$ | Program $C_e$ |
|---|---|---|
| ```i:=0;``` | ```i:=0;``` | ```x:=x+3;``` |
| ```while(i<n) {``` | ```k:=0;``` | ```n:=n+6;``` |
| ```    x:=x+5*i;``` | ```while(i<n) {``` | ```x:=n+2*i;``` |
| ```    i:=i+1;``` | ```    x:=x+k;``` | |
| ```}``` | ```    i:=i+1;``` | |
| | ```    k:=k+5;``` | |
| | ```}``` | |

**Fig. 6.** Strength Reduction

***Induction Variable Elimination*** The source program $C_1$ in Fig. 7 is transformed to a target program $C_2$ in Fig. 7 by induction variable elimination. We assume that the environment of $C_1$ does not modify $i$ and $k$ and the environment of $C_2$ does not modify $k$. We verified the correctness of this transformation. We constructed a program $C_e$ in Fig. 7 and verified that $C_2||C_e$ refines $C_1||C_e$.

| Source Program $C_1$ | Target Program $C_2$ | Program $C_e$ |
|---|---|---|
| ```
i:=0;
k:=0;
while(i<n) {
    x:=x+k;
    i:=i+1;
    k:=k+5;
}
``` | ```
k:=0;
while(k<5*n) {
    x:=x+k;
    k:=k+5;
}
``` | ```
x:=x+3;
n:=n+6;
x:=n+2*x;
``` |

**Fig. 7.** Induction Variable Elimination

***Dead Code Elimination*** The source program $C_1$ in Fig. 8 is transformed to the target program $C_2$ in Fig. 8 by dead code elimination. We assume that the environments of $C_1$ and $C_2$ does not modify the values of $x, y$ and $z$. We constructed a program $C_e$ in Fig. 8 and found that $C_2 || C_e$ does not refine $C_1 || C_e$.

| Source Program $C_1$ | Target Program $C_2$ | Program $C_e$ |
|---|---|---|
| ```
x:=1;
y:=6;
z:=4;
y:=10;
z:=3;
``` | ```
x:=1;
y:=10;
z:=3;
``` | ```
a:=y+z;
x:=1*x;
a:=2*x+a;
``` |

**Fig. 8.** Dead Code Elimination

***Constant Folding and Constant Propagation*** The source program $C_1$ in Fig. 9 is transformed to a target program $C_2$ in Fig. 9 by constant folding and constant propagation. Both $C_1$ and $C_2$ are supposed to run in an environment that does not modify the values of $x$ and $y$. We constructed a program $C_e$ in Fig. 9 and verified that $C_2 || C_e$ refines $C_1 || C_e$.

| Source Program $C_1$ | Target Program $C_2$ | Program $C_e$ |
|---|---|---|
| ```
x:=100;
y:=x+6;
if (y<200) {
    z:=x+y;
}
``` | ```
x:=100;
y:=106;
if (true) {
    z:=206;
}
``` | ```
z:=x+3;
a:=x;
z:=z+2;
``` |

**Fig. 9.** Constant Folding and Constant Propagation

***If Optimization (1)*** The source program $C_1$ in Fig. 10 is transformed to the target program $C_2$ in Fig. 10 by if optimization. Both $C_1$ and $C_2$ are supposed to run in an environment that does not change the values of $x$ and $n$. We constructed a program $C_e$ in Fig. 10 and verified that $C_2 || C_e$ refines $C_1 || C_e$.



```
Source Program C₁    Target Program C₂    Program Cₑ
if (x>n) {           if (x>n) {           y:=x+y;
    y:=y+1;              y:=y+1;          z:=y+2;
    if (x>n) {           z:=z+1;          b:=2*x+z;
        z:=z+1;          x:=y+z;          z:=z+b;
    }                }
    x:=y+z;
}
```

**Fig. 10.** If Optimization (1)

*If Optimization (2)* The source program $C_1$ in Fig. 11 is transformed to the target program $C_2$ in Fig. 11 by if optimization. Both $C_1$ and $C_2$ are supposed to run in an environment that does not change the values of $x$ and $n$. We constructed a program $C_e$ in Fig. 11 and verified that $C_2||C_e$ refines $C_1||C_e$.

```
Source Program C₁    Target Program C₂    Program Cₑ
if (x>n) {           if (x>n) {           y:=x+y;
    y:=y+1;              y:=y+1;          z:=y+2;
}                        z:=z+1;          b:=2*x+z;
if (x>n) {           }                    z:=z+b;
    z:=z+1;              x:=y+z;
}
x:=y+z;
```

**Fig. 11.** If Optimization (2)

*Cross Jumping* The source program $C_1$ in Fig. 12 is transformed to the target program $C_2$ in Fig. 12 by cross jumping. We allow the environments of both $C_1$ and $C_2$ to have arbitrary behaviours. We constructed a program $C_e$ in Fig. 12 and verified that $C_2||C_e$ refines $C_1||C_e$.

```
Source Program C₁    Target Program C₂    Program Cₑ
if (x>0) {           if (x>0) {           x:=a+b;
    a:=a+5;              a:=a+5;          b:=a+2;
    if (y<0) {       }                    c:=5*a+b;
        b:=0;        else {               y:=x-5;
    }                    b:=b+4;          b:=x+c;
}                    }
else {               if (y<0) {
    b:=b+4;              b:=0;
    if (y<0) {       }
        b:=0;
    }
}
```

**Fig. 12.** Cross Jumping



**Branch Elimination** The source program $C_1$ in Fig. 13 is transformed to the target program $C_2$ in Fig. 13 by branch elimination. Both $C_1$ and $C_2$ are supposed to run in an environment that does not modify the value of $a$. We constructed a program $C_e$ in Fig. 13 and verified that $C_2 || C_e$ refines $C_1 || C_e$.

```
Source Program C₁          Target Program C₂          Program Cₑ
while  (i<n) {              if (a<5) {                 i:=a+2;
    if  (a<5) {                while  (i<n) {          b:=a+b;
        b:=b+5;                    b:=b+5;             c:=2*b-i;
    }                              i:=i+1;             a:=1*a;
    else {                     }                       b:=i+3*c;
        b:=2*b;            }
    }                      else {
    i:=i+1;                    while  (i<n) {
}                                  b:=2*b;
                                   i:=i+1;
                               }
                           }
```

**Fig. 13.** Branch Elimination

**Loop Unrolling** The source program $C_1$ in Fig. 14 is transformed to the target program $C_2$ in Fig. 14 by loop unrolling. Both $C_1$ and $C_2$ to are supposed to run in an environment that does not modify the value of $i$. We constructed a program $C_e$ in Fig. 14 and verified that $C_2 || C_e$ refines $C_1 || C_e$.

```
Source Program C₁          Target Program C₂          Program Cₑ
i:=0;                      i:=0;                      x:=x+2;
while(i<2*n) {             while(i<2*n) {             a:=x+4;
    x:=x+1;                    x:=x+1;                x:=2*x;
    i:=i+1;                    x:=x+1;
}                              i:=i+2;
                           }
```

**Fig. 14.** Loop Unrolling